\documentstyle[aps,pre,preprint,eqsecnum]{revtex}
\begin{document}
\renewcommand{\thesection}{\arabic{section}}
\draft
\title{Three-point correlation function of a scalar mixed by an almost smooth
random velocity field}
\author{E. Balkovsky$^a$, G. Falkovich$^a$ and V. Lebedev$^{a,b}$}
\address{$^a$ Physics of Complex Systems, Weizmann Institute of Science,
Rehovot 76100, Israel \\ $^b$  Landau Institute for Theor. Physics, 
Moscow, Kosygina 2, 117940, Russia}
\date{\today}
\maketitle

\begin{abstract}
We demonstrate that if the exponent $\gamma$ that measures non-smoothness
of the velocity field is small then the isotropic zero modes of the scalar's 
triple correlation function have the scaling exponents proportional to
$\sqrt{\gamma}$. Therefore, zero modes are subleading with 
respect to the forced solution that has normal scaling with the exponent $\gamma$.
\end{abstract}

\section*{Introduction}

Kraichnan's model of passive scalar advection by white-in-time velocity field
\cite{68Kra-a} has became a paradigm within which analytic theory of
anomalous scaling in turbulence starts to appear
\cite{94SS,95GK,95SS,95CFKLb,95CF,96BGK,Yak,Pum,BG}. 
The instrument is the perturbation theory around three limiting cases where
scalar statistics is Gaussian: i) infinite space dimensionality 
\cite{68Kra-a,95CFKLb,95CF,BG}, ii) extremely irregular velocity field
which corresponds to a smooth scalar field \cite{95GK,96BGK,Pum}, and
iii) smooth velocity field (the Batchelor-Kraichnan limit) 
\cite{68Kra-a,94SS,95SS,96SS,Bat,74Kra-a,95CFKLa,95BCKL}.
The perturbation theory is regular in the first two cases and the sets
of the exponents obtained agree when the limits intersect \cite{95CF,96BGK}.
The perturbation theory around the Batchelor-Kraichnan limit 
is singular \cite{95SS}, only dipole part of three-point correlation
function has been found so far \cite{96PSS}. In this paper, motivated
by \cite{95SS,96PSS}, we find the isotropic part of the triple
correlation function and show that the leading term has a normal scaling at small scales.

It is instructive to discuss first the physics involved to understand
the significant difference between the first two limits on the one hand
and the third limit on the other. Since the scalar field at any point
is the superposition of fields brought from $d$ directions then it
follows from a central limit theorem that scalar's statistics approaches
Gaussian when space dimensionality $d$ increases. In the case ii), an
irregular velocity field acts like Brownian motion so that turbulent
diffusion is much like linear diffusion: statistics is Gaussian provided
the input is Gaussian. What is general in both limits is that
Gaussianity is rather uniform over the scales, the degree of Gaussianity
(say, flatness) is independent of the ratio $r/L$ where $r$ is a typical
distance in the correlation function and $L$ is an input scale.
Quite contrary, $\ln(L/r)$ is the parameter of Gaussianity in
the Batchelor-Kraichnan limit \cite{95CFKLa,FalLeb} so that statistics
is getting Gaussian at small scales whatever the input statistics.
The mechanism of Gaussianity is temporal rather than spatial in this case:
since the stretching is exponential in a smooth velocity field 
then the cascade time grows logarithmically as the scale decreases.
This is opposite to what one expects from intermittency and anomalous
scaling (anticipated beyond the Gaussian limits):
the degree of non-Gaussianity has to grow downscales. Already that
simple reasoning shows that the way from the Batchelor-Kraichnan limit
towards an anomalous scaling at non-smooth velocity field is not to be
easy. The formal reason for this perturbation theory to be singular is
that, at the limit, the many-point correlation functions have singularity
(smeared by molecular diffusion only) at the collinear geometry 
-- smooth velocity provides for homothetic transformation that does not
break collinearity \cite{95BCKL}. Even weak non-smoothness of the velocity
smears the singularity i.e. strongly influences the solution in the
narrow region near collinearity; such a situation calls for a boundary
layer approach introduced into this problem by Shraiman and Siggia 
\cite{95SS,96SS}.

\section{General Relations}

The triple correlation function of the scalar
$F({\bf r}_1,{\bf r}_2,{\bf r}_3 )$ advected by white-in-time velocity
field satisfies the closed balance equation \cite{94SS}
\begin{equation}
(\hat{\cal L}+\hat{\cal L}_d)F_3=-\chi_3\,.
\label{eq} \end{equation}
Here, $\chi_3({\bf r}_1,{\bf r}_2,{\bf r}_3 )$ is the triple correlation
function of the (non-Gaussian) pumping force. Actually it depends on differences
${\bf r}_{ij}={\bf r}_i-{\bf r}_j$. If $|{\bf r}_{ij}|\ll L$ (where
$L$ is the pumping length) then $\chi_3\simeq P_3$ where $P_3$ is the
third-order flux. At growing $|{\bf r}_{ij}|$ the function $\chi_3$
tends to zero on distances larger than $L$. The quantity $\hat{\cal L}_d$
in (\ref{eq}) is the operator of molecular diffusion
$\hat{\cal L}_d=\kappa(\nabla_1^2+\nabla_2^2+\nabla_3^2)$
($\kappa$ is the diffusion coefficient), and
\begin{equation} 
\hat{\cal L}=-(1/2)\sum_{i,j=1}^3 
{\cal K}_{\alpha\beta}({\bf r}_{ij})\nabla_i^\alpha\nabla_j^\beta
\label{eq1} \end{equation}
is the operator of turbulent diffusion. Here 
\begin{equation}
{\cal K}^{\alpha\beta}({\bf r})=
Dr^{-\gamma}\left[(r^2\delta^{\alpha\beta}-r^\alpha r^\beta)
+\frac{d-1}{2-\gamma}r^2\delta^{\alpha\beta}\right]
\label{eq2} \end{equation}
is the eddy diffusivity related to the velocity pair correlation function.
Parameter $\gamma$ in (\ref{eq2}) is a measure of velocity non-smoothness, $0\leq\gamma\leq2$. 

At $\gamma=0$, the operator $\hat{\cal L}$ is singular for collinear geometry --- see
(\ref{h1}) below. That leads to an angular singularity in the correlation
functions which is smoothed only by diffusion which is therefore relevant at all
scales \cite{95BCKL}. Contrary, at $\gamma>0$
the operator $\hat{\cal L}$ is not singular at the collinear geometry and therefore the
angular singularity is absent as was pointed out in \cite{96PSS}.
Therefore we can omit the diffusive term $\hat{\cal L}_d$ in (\ref{eq}) in comparison with
$\hat{\cal L}$. This is possible as long as $\kappa\ll\gamma Dr^2$.

In the following we believe $\chi_3$ to be an isotropic function of 
${\bf r}_{ij}$ what dictates the symmetry of the solution of (\ref{eq}).
In this case $F_3$ can be treated as a function of three distances
$r_{12}$, $r_{13}$ and $r_{23}$ only. Then the operator $\hat{\cal L}$ (\ref{eq1})
can also be rewritten in terms of the separations \cite{95CFKLa}
\begin{equation} 
\hat{\cal L}=\frac{D(d-1)}{2-\gamma}
\sum\limits_{i>j}r_{ij}^{1-d}\partial_{ij}
r_{ij}^{1+d-\gamma}\partial_{ij}+\dots
\label{eq3} \end{equation}
where dots stand for the terms with cross derivatives $\partial_{ij}\partial_{kl}$. 
Since $\chi_3=P_3$
at $r_{ij}\ll L$ we can easily find a solution of (\ref{eq}) in the 
region $r_{ij}\ll L$ (cf. \cite{95CFKLa}). Using (\ref{eq3}) we get
\begin{equation}
F_{\rm forc}=\frac{(2-\gamma)P_3 L^\gamma}{3D(d-1)d\gamma}
\left[C-\left(\frac{r_{12}}{L}\right)^\gamma
-\left(\frac{r_{13}}{L}\right)^\gamma
-\left(\frac{r_{23}}{L}\right)^\gamma\right]\,,
\label{eq4} \end{equation}
which we will call the forced solution. Here $C$ is a constant of the order unity. 
That solution satisfies the equation $r_{ij}\ll L$ but not necessarily matching 
conditions at $r_{ij}\gtrsim L$. The solution of (\ref{eq}) that satisfies the condition
can be written at $r_{ij}\ll L$ as follows:
\begin{equation}
F=F_{\rm forc}+Z_0\,,
\label{eq5} \end{equation}
where $Z_0$ is a zero mode of the operator of the turbulent diffusion: $\hat{\cal L} Z_0=0$. 
We examine here different solutions of the equation $\hat{\cal L} Z=0$. Whether the given
mode $Z$ contributes the correlation function has to be determined from the matching
at $r_{ij}\sim L$ which is beyond the scope of our paper. Note that we consider isotropic 
zero modes while dipole zero modes for the anisotropic problem with an imposed mean 
gradient were treated in \cite{96PSS}.

It is convenient to introduce instead of $r_{ij}$ the following 
set of variables
\begin{eqnarray} &&
x_1=\frac{r_{13}}{r_{12}}\cos\theta,\quad
x_2=\frac{r_{13}}{r_{12}}\sin\theta,\quad
s=r_{12}r_{13}\sin\theta\,,
\label{qg1} \end{eqnarray}
where $\theta$ is the angle between ${\bf r}_{12}$ and ${\bf r}_{13}$ and
$-\infty<x_1<\infty$, $0<x_2<\infty$, $0<s<\infty$. Note that the 
variable $s$ (which is the doubled area of the triangle) is the only dimensional
parameter among $s$, $x_1$, $x_2$. The expressions inverse to (\ref{qg1}) are
\begin{eqnarray} &&
r_{12}=\sqrt{\frac{s}{x_2}}\,, \quad
r_{13}=\sqrt{\frac{s}{x_2}(x_1^2+x_2^2)}\,,
\nonumber \\ &&
r_{23}=\sqrt{\frac{s}{x_2}\big[(x_1-1)^2+x_2^2\big]}\,, \quad
\cos\theta=\frac{x_1}{\sqrt{x_1^2+x_2^2}}\,.
\label{qg3} \end{eqnarray}
The operator $\hat{\cal L}$ and both correlation functions $\chi_3$
and $F_3$ should be invariant under permutation of points ${\bf r}_1$,
${\bf r}_2$ and ${\bf r}_3$, that is under permutations of $r_{12}$,
$r_{13}$ and $r_{23}$. In terms of the variable $z=x_1+ix_2$ these 
transformations can be written as follows
\begin{eqnarray} &&
1\leftrightarrow 2:\quad z\rightarrow 1-z^*,\quad 
2\leftrightarrow 3:\quad z\rightarrow \frac{1}{z^*},\quad 
1\leftrightarrow 3:\quad z\rightarrow 1+\frac{1}{z^*-1},
\label{g22} \\ &&
1\rightarrow 2\rightarrow 3:\quad z\rightarrow \frac{1}{1-z},\quad 
1\rightarrow 3\rightarrow 2:\quad z\rightarrow 1-\frac{1}{z},
\label{g2} \end{eqnarray}
where $z^*$ is complex conjugated to $z$. The doubled area of the triangle $s$ is 
obviously invariant under the transformations.

\section{The case $\gamma=0$}

Here, we start with the case $\gamma=0$. Below we treat the dimensionality $d=2$. 
The results can be generalized for an arbitrary dimensionality $d$.
The operator (\ref{eq1}) for $d=2$ and $\gamma=0$ is rewritten 
in terms of the variables (\ref{qg1}) as follows
\begin{equation}
\hat{\cal L}_0=2Dx_2^2(\partial_1^2+\partial_2^2)\,.
\label{h1} \end{equation}
Let us stress that derivatives with respect to $s$ are absent in $\hat{\cal L}_0$.
Then a solution of the equation $\hat{\cal L}_0F_3=-\chi_3$ can be written as
\begin{equation}
F_3(s,x_1,x_2)=\frac{1}{8\pi D}\int_{-\infty}^{+\infty}\int_{0}^{\infty}
\frac{dx_1'dx_2'}{x_2'^2}\ln\left[\frac{(x_1'-x_1)^2+(x_2'+x_2)^2}
{(x_1'-x_1)^2+(x_2'-x_2)^2}\right]\chi_3(s,x_1',x_2')\,,
\label{eq8}\end{equation}
where we used the explicit expression for the resolvent of 
Laplacian (cf. \cite{95BCKL}). 

We are interested in the behavior of the correlation functions at 
$r_{ij}\ll L$ which is not sensitive to the particular form  of the
pumping $\chi_3$. Thus we can choose any convenient form of $\chi_3$
supplying the convergence of the integral in (\ref{eq8}). We get
the following function
\begin{eqnarray}&&
\chi_3=\frac{P_3}{1+(r_{12}^2+r_{13}^2+r_{23}^2)/L^2}.
\label{h32}\end{eqnarray}
Then the correlation function is given by 
\begin{eqnarray}&&
F_3=\frac{P_3}{8\pi D}\int\limits_{-\infty}^{+\infty}\int\limits_{0}^{\infty}
\frac{dx_1'dx_2'}{x_2'^2}
\frac{x_2'}{x_2'+2sL^{-2}(x_1'^2+x_2'^2+1-x_1')}\ln\left[
\frac{(x_1'-x_1)^2+(x_2'+x_2)^2}{(x_1'-x_1)^2+(x_2'-x_2)^2}\right].
\label{h33}\end{eqnarray}
Performing the integration over $x_1'$ one obtains
\begin{eqnarray} &&
F_3=\frac{L^2P_3}{16sD}\int_{0}^{\infty}\frac{dt}{tw}\ln\left[
\frac{(x_2(1+t)+w)^2+(x_1-1/2)^2}{(x_2|1-t|+w)^2+(x_1-1/2)^2}\right],
\label{h34}\end{eqnarray}
where $w=\sqrt{3/4+x_2^2t^2+L^2x_2t/(2s)}$ (cf. \cite{95BCKL}).
The asymptotics of $F_3$ at $r_{ij}\ll L$ is 
\begin{eqnarray}&&
F_3=\frac{P_3}{2D}\ln\left[\frac{2L^2}{r_{12}^2+r_{13}^2+r_{23}^2
+2\sqrt3 s}\right]+{\rm const}\,.
\label{h35}\end{eqnarray}

The expression (\ref{h35}) can be rewritten as
\begin{equation}
F_3=\frac{P_3}{3D}\ln\frac{L^3}{r_{12}r_{13}r_{23}}
+\frac{P_3}{6D}\ln\frac{(x_1^2+x_2^2)[(x_1-1)^2+x_2^2]}
{[(x_1-1/2)^2+(x_2+\sqrt3/2)^2]^3}+{\rm const}\,.
\label{h36}\end{equation}
Here, the first term is the forced solution $F_{\rm forc}$ which can be obtained from
(\ref{eq4}) at $\gamma\to0$, $d=2$, $C=3$. The second term in (\ref{h36})
is the zero mode $Z_0$ which has logarithmic singularities in the points $z=0$, 
$z=1$ and $z=\infty$ that is where one of $r_{ij}$ tends to zero. 
Nevertheless the function $F_3$ has no singularity where one of $r_{ij}$ 
tends to zero as it is obvious from the expression (\ref{h35}). Further, as 
follows from (\ref{h36}) the zero mode $Z_0$ has the linear term in the expansion
over $x_2$. Remembering (\ref{qg1}) we conclude that at small angles 
$\theta$ there is the term proportional to $|\theta|$ in $F_3$. That is
just the angular singularity noted above. Repeat again that the singularity
is smoothed only by diffusion \cite{95BCKL}.

We see that the zero mode $Z_0$ at $\gamma=0$ does not depend on the dimensional 
parameter $s$. Besides, any function of $s$ is a zero mode of the operator
$\hat{\cal L}_0$ since it does not contain the derivative over $s$ what is a remarkable 
property of the case $d=2$ and $\gamma=0$. Nevertheless all the zero modes do not 
contribute to $Z_0$.

\section{Asymptotics at small angles} 

Because of the scaling properties of $\hat{\cal L}$,  it is possible to seek
the zero mode in the form $Z\propto s^\Delta$. It is convenient to 
introduce the following function
\begin{eqnarray} &&
Z=\left(\frac{s}{x_2}\right)^\Delta
\left\{1+(x_1^2+x_2^2)^{\Delta}+[(1-x_1)^2+x_2^2]^{\Delta}
\right\}X(x_1,x_2)\,,
\label{gg9} \end{eqnarray}
The function $X(x_1,x_2)$ should be invariant under all transformations
(\ref{g22},\ref{g2}) and have no angular singularities since the
proportionality coefficient between $Z$ and $X$ is equal to
$r_{12}^{2\Delta}+r_{13}^{2\Delta}+r_{23}^{2\Delta}$ as follows
from (\ref{qg3}).

The equation $\hat{\cal L} Z=0$ can be rewritten as $\hat{\cal L}_X X=0$ where 
$\hat{\cal L}_X$ is a differential operator of the second order over
$\partial_1\equiv\partial/\partial x_1$ and
$\partial_2\equiv\partial/\partial x_2$. Coefficients at the
derivatives are the functions of $x_1,x_2,\Delta$ which can be found
from (\ref{eq},\ref{qg3}). The functions are quite complicated.
Fortunately only particular parts of the operator $\hat{\cal L}_X$ will be
needed for us. 

At $\gamma=0$ the operator $\hat{\cal L}_X$ is determined by (\ref{h1}). The
operator tends to zero at $x_2\to0$. Therefore at small $x_2$ 
besides (\ref{h1}) we should take into account also the residue.
The term leading at small $x_2$  can be written as
\begin{eqnarray} &&  
\hat{\cal L}_X\propto \hat{\cal L}_2=
[2x_2^2+c_0(x_1)]\partial_2^2 
-4\Delta x_2\partial_2+2\Delta(\Delta+1)\,,
\label{q4} \\ &&
c_0(x)=-\frac{3\gamma}{4}
(1-x)x\big[x\ln|x|+(1-x)\ln|1-x|\big]\,.
\label{q3}  \end{eqnarray}
Note that $c_0>0$. The expression (\ref{q3}) is correct if 
$x_2\ll|x_1|,|x_1-1|$; $|x_1|,|x_1-1|\gg\exp(-1/\gamma)$;
$|x_1|,|x_1-1|\ll\exp(1/\gamma)$. The asymptotic behavior of a solution of 
$\hat{\cal L}_2 X=0$ at $x_2\ll\sqrt{c_0}$ is
\begin{equation}
X=A_1(x_1) +A_2(x_1) x_2 \,,
\label{gg2} \end{equation}
where $A_1(x)$ and $A_2(x)$ are arbitrary functions.
The analyticity of $X$ at small angles excludes the second
term in (\ref{gg2}) since it supplies the contribution
to $X$ which behaves $\propto|\vartheta|$.  
The equation $\hat{\cal L}_2 X=0$ can be solved explicitly, a solution 
having the asymptotics (\ref{gg2}) with $A_2=0$ is
\begin{equation} 
X_0=A_1(x_1)F\left(-\frac{1+\Delta}{2},-\frac{\Delta}{2};
\frac{1}{2};-\frac{2x_2^2}{c_0(x_1)}\right).
\label{q5} \end{equation}
Here, $F(\alpha,\beta;\gamma;z)$ is the hypergeometric function.

The expression (\ref{q5}) gives the behavior of the zero mode in the 
vicinity of the boundary layer $x_2\sim \sqrt{c_0}$. If we are interested 
in the behavior of the zero mode outside the boundary layer then 
it is more convenient to return to $Z$ since $\hat{\cal L}$ can be
approximated as (\ref{h1}) and consequently $Z$ is a harmonic function there. 
The asymptotics of (\ref{q5}) valid at $x_2\gg\sqrt{c_0}$ gives 
\begin{equation}
Z\propto (\Delta+1)\cos(\pi\Delta/2)
-x_2 \sin(\pi\Delta/2)
\left(\frac{2}{c_0(x_1)}\right)^{1/2}\,.
\label{h5} \end{equation}
The behavior (\ref{h5}) occurs outside the boundary layer but at small
$x_2$. Since we are interested in small $\Delta$ we can suggest $\Delta\ll1$.
Thus we come to the following problem: find the harmonic function
$Z(x_1,x_2)$ in the upper half-plane $x_2>0$ at the boundary condition  
\begin{equation}
\frac{\sqrt{2c_0}}{\pi\Delta}
\partial_2 Z+Z=0,
\label{h8} \end{equation}
which is imposed on the function $Z$ at $x_2=0$ since at small $\gamma$ the
width of the boundary layer is negligible.  

Let us show that there is no zero mode with $\Delta\ll\sqrt\gamma$.
A harmonic function $Z(x_1,x_2)$ inside the region can be presented as an integral
of its normal derivative $\partial Z/\partial n$ along the contour which is the
boundary of the region: 
\begin{equation}
Z(z)=\frac{1}{\pi}\oint |dt|\frac{\partial Z(t)}{\partial n}\ln|z-t|\,,
\label{cont} \end{equation}
where $z=x_1+ix_2$ and $t$ is the complex variable going along the contour.
Let us consider the contour consisting of the semi-circles going around the singular
points $0,1,\infty$ and the parts near the real axis 
(outside the boundary layer but at small $x_2$) that link the semicircles.
If $\Delta\ll\sqrt\gamma$ then the condition (\ref{h8}) tells us that in this 
case only contributions to (\ref{cont}) from the semicirles 
are relevant since the contributions from the parts 
of the real axis are negligible. Separate consideration
of the vicinities of the singular points $0,1,\infty$
(see below) shows that the logarithmic derivative of $Z$ has to be bounded
there. Thus the only possible contribution to the zero mode associated, say,
with the singular point $z=0$ is $\propto{\rm Re}\ln z=\ln\sqrt{x_1^2+x_2^2}$.
The complete zero mode should be symmetric under transformations
(\ref{g22}). Performing all the transformations to $\ln\sqrt{x_1^2+x_2^2}$ 
and summing the results we obtain zero. That means that the function
possessing the required symmetry does not exist. 

Another (equivalent) way of showing that there is no zero mode with $\Delta\ll\sqrt\gamma$
is to continue $Z$ to negative $x_2$ by $Z(x_1,x_2)=Z(x_1,-x_2)$. Becase in our case
we can believe $\partial_2 Z(x_1,x_2)=0$ the function should be harmonic in semicircles 
 surrounding the singular points $0,1,\infty$. Then one can use the properties of the 
analytical functions in the circles to exclude the existence
of zero modes with bounded logarithmic derivatives near the singular points.
Note the difference with the dipole case where such mode has been found \cite{96PSS}.

\section{Vicinities of Singular Points} 

Here, we describe the set of zero modes that do not have additional smallness of $\Delta$
relative to $\sqrt\gamma$ so that the whole boundary condition (\ref{h8}) is to be accounted.
Necessary information about the structure of the modes can be extracted from
the analysis of the vicinities of the singular points $z=0$, $z=1$ and $z=\infty$ where one
needs a separate consideration. Using the symmetry properties 
(\ref{g22},\ref{g2}) we can reduce the consideration to the vicinity 
of one of the points, say $z=1$. At $x_2\ll1$ and $|x_1-1|\ll1$
the operator $\hat{\cal L}_X$ acquires the following form
\begin{eqnarray} &&
\hat{\cal L}_X=\mu\left[\rho^2\partial_\rho^2+3\rho\partial_\rho
+3\partial_\varphi^2\right]+2\sin^2\varphi
\left[\rho^2\partial_\rho^2+\rho\partial_\rho
+\partial_\varphi^2\right]
\nonumber \\ &&
-4\Delta[\sin^2\varphi\rho\partial_\rho
+\cos\varphi\sin\varphi\partial_\varphi]+2\Delta(\Delta+1) \,,
\label{h15} \\ && 
x_1-1=\rho\cos\varphi\,, \quad
x_2=\rho\sin\varphi\,, \quad \mu=\frac{1}{2}(\rho^{-\gamma}-1)\,.
\end{eqnarray} 

In the exponentially narrow vicinity of the singular point where $\rho\ll\exp(-1/\gamma)$ 
one has $\mu\gg1$ and the equation $\hat{\cal L}_XX=0$ is reduced to
\begin{equation}
\left[\rho^2\partial_\rho^2+3\rho\partial_\rho
+3\partial_\varphi^2\right]X=0\,.
\label{hh15} \end{equation}
Solutions of the equation (\ref{hh15}) can be expanded into the
Fourier series over the angular harmonics
\begin{equation}
X_m\propto\sin(m\varphi)\rho^{\lambda_m}\,.
\label{h16} \end{equation}
Substituting (\ref{h16}) into (\ref{hh15}) we obtain
\begin{equation}
\lambda_m=-1+\sqrt{1+3m^2},
\label{h17} \end{equation}
where only nonnegative $\lambda_m$ are taken since $X$ should remain
finite at $\rho\to0$. Thus besides a constant corresponding to $m=0$ we 
obtain the next term corresponding to $m=1$ that behaves $\propto\rho$.  
Since inside the exponentially narrow vicinity a zero mode $X$ is a linear
combination of (\ref{h16}) we come to the conclusion that the matching 
condition on the boundary of the vicinity should be imposed on the
logarithmic derivative of $X$ (or $Z$) which remains constant there.

Now, let us consider the region $1\gg\rho\gg\exp(-1/\gamma)$ where
\begin{equation}
\mu=\frac{1}{2}\gamma\ln\frac{1}{\rho}\ll1.
\label{h18} \end{equation}
Here, we can consider separately small angles $\varphi\ll1$ where
$\hat{\cal L}_XX=0$ is reduced to
\begin{equation}
\left[(3\mu+2\varphi^2)\partial_\varphi^2
-4\Delta\varphi\partial_\varphi
+2\Delta(\Delta+1)\right]X=0 \,,
\label{h19} \end{equation}
what exactly corresponds to $\hat{\cal L}_2X=0$. Again an appropriate solution of (\ref{h19})
is as follows
\begin{equation}
X\propto F\left(-\frac{1+\Delta}{2},-\frac{\Delta}{2};
\frac{1}{2};-\frac{2\varphi^2}{3\mu}\right)\,,
\label{h20} \end{equation}
that gives the asymptotics
\begin{equation}
Z\propto 1-\frac{\pi\Delta}{\sqrt{6\mu}}\varphi\,,
\label{h21} \end{equation}
at $1\gg\varphi\gg\sqrt\mu$. Substituting now (\ref{h18}) we come to
the boundary condition 
\begin{equation}
\partial_\varphi Z=
-\frac{\pi\Delta}{\sqrt{3\gamma\ln(1/\rho)}} Z\,,
\label{h22} \end{equation}
imposed on the harmonic function $Z$ at $\varphi=0$. Note that (\ref{h22}) is nothing but
the limit of (\ref{h8}) at $\rho\ll1$. That boundary condition is simple enough and
permits explicit expression for $Z$ near the singularity. 

Let us represent $Z$ in the following form
\begin{equation}
Z={\rm Re}\left\{ f\left(\ln\frac{1}{\rho}+i\varphi\right)+
f\left(\ln\frac{1}{\rho}
+i\pi-i\varphi\right)\right\}
\label{h23} \end{equation}
which is obviously harmonic and invariant under $\varphi\to\pi-\varphi$. 
Taking into account $\ln(1/\rho)\gg1$ we obtain from (\ref{h23})
\begin{eqnarray} &&
Z(\rho,\varphi=0)=
2{\rm Re}\left\{f\left(\ln\frac{1}{\rho}\right)\right\}\,,
\label{h24} \\ &&
\frac{\partial Z}{\partial\varphi}(\rho,\varphi=0)
=\pi{\rm Re}\left\{f''\left(\ln\frac{1}{\rho}\right)\right\}.
\label{h25} \end{eqnarray}
Thus we see that $f(x)$ satisfies the following equation
\begin{equation}
f''(x)+\frac{2\Delta}{\sqrt{3\gamma x}}f(x)=0.
\label{h26} \end{equation}
Asymptotic behavior of the solution at $x\gg 1$ is 
\begin{equation}
f(x)=\exp\left[\pm\frac{4}{3}i\left(\frac{4\Delta^2}{3\gamma}
\right)^{1/4}x^{3/4}\right]\,.
\label{h27} \end{equation}
Expanding (\ref{h23}) with (\ref{h27}) over $\ln^{-1}(1/\rho)$ we obtain
\begin{equation}
Z\propto\cos\left[\frac{4}{3}\left(\frac{4\Delta^2}{3\gamma}\right)^{1/4}
\left(\ln\frac{1}{\rho}\right)^{3/4}+\phi_0\right]
\left\{1-\frac{\Delta}{\sqrt{3\gamma}}
\frac{\varphi(\pi-\varphi)}{\sqrt{\ln(1/\rho)}}\right\}\,,
\label{h28} \end{equation}
where $\phi_0$ is some phase. We can believe $|\phi_0|<\pi$, its
actual value has to be determined by the matching at $-\ln\rho\sim\gamma$. 

Symmetry requirement with respect to $x_1\rightarrow1-x_1$ leads to the condition 
$\partial_1 Z(1/2,x_2)=0$ which can be used as the quantization rule for the
zero modes having the asymptotics (\ref{h28}). They can be classified in accordance
with the number of zeros $n$ which the function $Z$ have where $x_1$ goes from
$1/2$ to $1-\exp(-\gamma)$. Using the expression (\ref{h28}) we conclude that
$\Delta_{min}=\alpha\sqrt\gamma$ and for $n\gg1$
\begin{equation}
\Delta_n=\beta\sqrt{\gamma}n^2\label{spectrum}\end{equation}
with yet unknown numerical factors $\alpha,\beta$ which are of order unity. 
Note that nonsymmetric zero modes $Z$ (with another values of $\alpha,\beta$) 
may exist yet they cannot contribute to $Z_0$.
For all the modes, the dependence $\Delta(\gamma)$ obtained here
has an infinite slope at zero which has been also observed in numerics \cite{chem}.
We conclude that the set of zero modes thus found at small $\gamma$ has exponents larger
than the exponent $\gamma$ of forced solution. Therefore, the isotropic
part of the triple correlation function is shown here to have a normal
scaling for sufficiently small $\gamma$. Since at $\gamma=2$ the lowest
zero mode has $\Delta=4$, it is likely that the scaling of the isotropic part of
the triple correlation function is normal for all $\gamma$.

\acknowledgements
We are grateful to M. Chertkov and B. Shraiman for useful discussions. 
The work was supported by the Israel Science Foundation (E.B.),
by the Minerva Center for Nonlinear Physics 
 and by the Cemach and Anna Oiserman Research Fund (G.F.).
V.L. is a Meyerhoff Visiting Professor at the Weizmann Institute.

\end{document}